\documentclass[10.5 pt,twocolumn]{article}

\usepackage{revista_poli-01}
\usepackage{multirow}

\usepackage{ctable} 

\titlerevista[An Extreme Rotating Black Hole in New Massive Gravity Theory]{An Extreme Rotating Black Hole \\ in New Massive Gravity Theory}%

\newcommand{\tituloingles}{ Agujero negro rotante extremo \\ en la teor\'ia \textit{New Massive Gravity}} 
\newcommand{\correoautor}{\footnotesize acena.andres@gmail.com; ericsson.lopez@epn.edu.ec} 

\authorrevista[2][Aceña Andrés]{Aceña Andrés$^{1,}$}
\authorrevista[1]{López Ericson}
\authorrevista[1][Llerena Mario]{Llerena Mario}

\institucionmail[1][Observatorio Astronómico de Quito, Unidad de Gravitación y Cosmología, Ecuador]{Escuela Politécnica Nacional,}{}
\institucionmail[2][Facultad de Ciencias Exactas y Naturales, CONICET, Mendoza, Argentina]{Universidad Nacional de Cuyo}{}
\newcommand{\approptoinn}[2]{\mathrel{\vcenter{
  \offinterlineskip\halign{\hfil$##$\cr
    #1\propto\cr\noalign{\kern2pt}#1\sim\cr\noalign{\kern-2pt}}}}}

\newcommand{\appropto}{\mathpalette\approptoinn\relax}
\renewcommand{\t}[1]{\tilde{#1}}

\newcommand{\ra}{\rightarrow}
\newcommand{\mR}{\mathbb{R}}

\nombrerevista{Revista Politécnica}
\mesrevista{Abril}
\volumenrevista{39}
\aniorevista{2017}
\numerorevista{1}
\sethdrevista

\makeatletter
\def\blfootnote{\xdef\@thefnmark{}\@footnotetext}
\makeatother

\begin{document}
\label{PrimeraPagina}

\fancypagestyle{plain}{
\fancyhead[R]{ }
\fancyhead[L]{ }
\fancyfoot[R]{ }
\fancyfoot[L]{{\rule[1.5cm]{5.2cm}{.4pt} \\[-1.5cm]\scriptsize \correoautor} }
\renewcommand{\headrulewidth}{0.5pt}
}
\pagestyle{fancy}

\twocolumn[
\begin{@twocolumnfalse}

\maketitlerevista

\renewcommand\abstractname{\vspace{-2\baselineskip}}

\hspace*{0.05\linewidth}\begin{minipage}{0.9\linewidth}
\begin{resumen}
\textbf{Abstract:} New Massive Gravity is an alternative theory to General Relativity that is used to describe the gravitational field in a (2+1)-dimensional spacetime. Black hole solutions have been found in this theory, in particular an asymptotically \mbox{anti-de Sitter} rotating black hole. We analyse some features of this solution as its event horizon, black hole area and distance to the horizon, specially in the rotating extreme case, showing that they have shared features with extreme black holes in 4-dimensional General relativity. This limit case is interesting in the search of geometric inequalities as the ones found for the Kerr black hole in (3+1)-General Relativity.

\keywordsrevista{New Massive Gravity, rotating black hole, extreme case}
\end{resumen}
\vspace*{-0.3cm}
\begin{center}
\LARGE\bf \tituloingles
\end{center} 
\vspace*{-0.3cm}
\begin{abstract}
\textbf{Resumen:} La teoría conocida como \textit{New Massive Gravity} es una teoría alternativa a la Relatividad General y describe el campo gravitacional en un espacio-tiempo (2+1)-dimensional. Dentro de esta teoría, se han encontrado soluciones de agujeros negros, en particular, se han encontrado soluciones de agujeros negros rotantes asintóticamente \mbox{anti-de Sitter}. Se analizan algunas propiedades de estas soluciones como su horizonte de eventos, el área del agujero negro y la distancia al horizonte, especialmente en el caso rotante extremo, mostrando que comparten caracter\'isticas con agujeros negros extremos en Relatividad general en 4 dimensiones. Este caso límite es interesante para poder hallar desigualdades geométricas como las que se han hallado para el agujero negro de Kerr en el contexto de la Relatividad General en dimensión (3+1). 

\palabrasclavesrevista{New Massive Gravity, agujero negro rotante, caso extremo}

\end{abstract}
\end{minipage}
\end{@twocolumnfalse}
]

\section{{INTRODUCTION}}

General Relativity is the theory used to describe gravitational phenomena, it provides impeccable agreement with observations. The Universe is modelled as a (3+1)-dimensional manifold (3 spacial dimensions and 1 time dimension). In this theory, the quanta associated with gravitational waves, i.e. gravitons, are massless particles with two independent polarization states, of helicity $\pm$2 (\cite{Berg}).
In order to have massive particles within the gravitational theory, alternative theories are needed and one should break one of the underlying assumptions behind the theory (\cite{Claudia}). \\

On the other hand, if we consider a (2+1)-dimensional manifold, i.e., (2+1)-General Relativity, it has been proven there are no gravitational waves in the classical theory (\cite{Carlip3}). Also, it has no Newtonian limit and no propagating degrees of freedom (\cite{Carlip}). But, in 1992, Ba\~nados, Teitelboim, and Zanelli (BTZ), to great surprise, showed that in (2+1)-dimensional gravity there is a black hole solution (\cite{Banados}). \\

This solution is called the BTZ black hole, it has an event horizon and (in the rotating case) an inner horizon. This black hole is asymptotically \mbox{anti-de Sitter}, and has no curvature singularity at the origin, so it is different from the known solutions in (3+1)-dimensions as the Schwarzschild and Kerr black holes, which are asymptotically flat (\cite{Carlip}).\\

The rotating BTZ metric is presented in \cite{Banados} and it is given by Equation \eqref{BTZ1}
\begin{equation}
ds^2=-f^2 dt^2+f^{-2}dr^2+r^2\left(d\phi+N^{\phi}dt\right)^2
\label{BTZ1}
\end{equation}
where $f=\left(-M+\frac{r^2}{l^2}+\frac{J^2}{4r^2}\right)^{\frac{1}{2}}$ and $
N^{\phi}=-\frac{J}{2r^2}$ with $|J|\leq Ml$, where $M$ and $J$ are the 
mass and angular momentum that are defined by the asymptotic symmetries at spatial infinity. The parameter $l$ is related with the cosmological constant as $\Lambda=-1/l^2$. This metric is stationary and axially symmetric, with Killing vectors $\partial_t$ and $\partial_{\phi}$, and generically has no other symmetries (\cite{Carlip2}). \\

The event horizons of the BTZ black hole are located at $r_{\pm}=\sqrt{\frac{Ml^2}{2}\left\lbrace  1\pm \left[1-\left(\frac{J}{Ml}\right)^2\right]^{\frac{1}{2}}\right\rbrace }$ and there would be a naked singularity if $|J|>Ml$. Also, if $M=-1$ and $J=0$, it is the \mbox{anti-de Sitter} (AdS) space-time (\cite{Banados}).\\

In this context, an alternative theory to General Relativity is New Massive Gravity (NMG), which is a theory that describes gravity in a vacuum (2+1)-spacetime with a massive graviton (\cite{Berg}). Some black hole solutions have been found for this theory, in particular rotating solutions, which are interesting in the search of geometric inequalities as the ones found for the Kerr black hole in (3+1)-General Relativity.\\

\begin{figure}[t!]
\centering
\includegraphics[width=0.80\columnwidth]{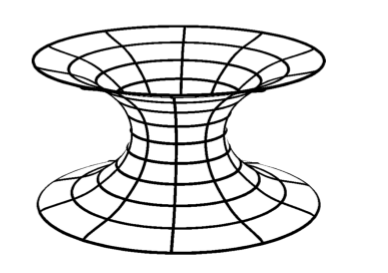}
\caption{Initial data with two asymptotically flat ends. Picture taken from (\cite{Dain2}.}
\label{kerrnoextr}
\end{figure}

\begin{figure}[t!]
\centering
\includegraphics[width=.8\columnwidth]{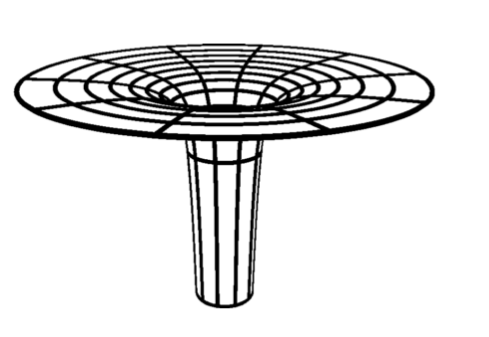}
\caption{The cylindrical end on extreme Kerr black hole initial data. Picture taken from 
(\cite{Dain2}).}
\label{kerrextr}
\end{figure}

In \cite{Dain} the known behaviour of the Kerr family is presented. Since the Kerr metric depends on two parameters, the mass $m$ and the angular momentum $J$, it was shown that it represents a black hole if and only if the following remarkably inequality holds $\sqrt{| J|}\leq m$. Otherwise the spacetime contains a naked singularity. So, there is an extreme Kerr black hole that reachs the equality $\sqrt{|J|}= m$ and it represents an object of optimal shape (\cite{Dain}). A graphical representation of the structure of a spatial slide of the Kerr black hole in the non-extreme case, which has two asymptotically flat ends, can be seen in Figure \ref{kerrnoextr}. For the extreme case, the corresponding representation is Figure \ref{kerrextr}, where it can be seen the asymptotically flat end, and also the cylindrical end, a typical feature of extreme initial data. \\ 

In this paper, we study a rotating black hole in NMG, considering specially its extreme limit with respect to the angular momentum parameter, as we are interested in its properties in the context of geometrical inequalities. The article is structured as follows. In section \ref{NMG}, we introduce briefly the NMG field equations. In section \ref{static} some static black hole solutions are presented while in section \ref{rotating} we present the rotating black hole solution that we analyse in this paper. In section \ref{analysis} we perform our analysis of the non-extreme and extreme cases, in particular their event horizons, area of some surfaces and the distance to their horizons. Finally, in section \ref{conclusions} we discuss the conclusions of our analysis.

\section{{NEW MASSIVE GRAVITY}}\label{NMG}

As we said before, in General Relativity, if we consider a (2+1)-spacetime we will find that there are no propagating degrees of freedom for a massless graviton (\cite{Claudia}). If we want to include a massive graviton that propagates degrees of freedom in the theory, we need an alternative theory. NMG is one of the alternatives (\cite{Berg}).\\

The action for NMG is $I=\frac{1}{16\pi G}\int d^3 x\sqrt{-g}\left(R-2\lambda-\frac{K}{m^2}\right)$, where $K=R_{\mu\nu}R^{\mu\nu}-\frac{3}{8}R^2$, and $G$ is the equivalent gravitational constant in a (2+1)-spacetime, while $m$ and $\lambda$ are parameters related with the cosmological constant (\cite{Berg, Oliva}). If the scalar $K$ is equal to zero, then the action of General Relativity is obtained, or if $m^2\rightarrow \infty$.\\

So, as shown in \cite{Oliva}, the vacuum field equations in this alternative theory are of fourth order and read
\begin{equation}
\label{fieldeq}
G_{\mu\nu}+\lambda g_{\mu\nu}-\dfrac{1}{2m^2}K_{\mu\nu}=0,
\end{equation}
where $K_{\mu\nu}$ is a tensor defined as Equation \eqref{kdefined}
\begin{equation}
\begin{array}{rl}
K_{\mu\nu}=&2\nabla_\rho \nabla^\rho R_{\mu\nu}-\tfrac{1}{2} \left( \nabla_{\mu} \nabla_{\nu} R +g_{\mu\nu}\nabla_\rho \nabla^\rho R\right) \\
		   &-8R_{\mu\rho}R^{\rho}\,_{\nu}+\tfrac{9}{2}R R_{\mu\nu} + g_{\mu\nu}\left(3R^{\rho\lambda}R_{\rho\lambda}-\tfrac{13}{8}R^2\right),
\end{array}
\label{kdefined}
\end{equation}
and $K=g^{\mu\nu}K_{\mu\nu}$. Black hole solutions have been found for the field equations in Equation \eqref{fieldeq}, we present some of them in the next section.

\section{{STATIC BLACK HOLE FAMILIES}}\label{static}

From \cite{Berg}, NMG theory generically admits solutions of constant curvature given by  
$R^{\mu\nu}\,_{\alpha\beta}=\Lambda\delta^{\mu\nu}_{\alpha\beta}$ with two different values of the parameter $\Lambda$, determined by $\Lambda_{\pm}=2m(m\pm \sqrt{m^2 -\lambda})$. This means that at the special case defined by $m^2=\lambda$ the theory possesses a unique maximally symmetric solution of constant curvature given by $\Lambda=\Lambda_{+}=\Lambda_{-}=2m^2=2\lambda$. In that case, the theory admits the following Euclidean metric as an exact solution (\cite{Oliva})
\begin{equation}
\label{metricgen}
ds^2=(-\Lambda r^2+br-\mu)d\psi^2 +\dfrac{dr^2}{-\Lambda r^2+br-\mu}+r^2 d\varphi^2,
\end{equation}
where $b$ and $\mu$ are integration constants, and $\Lambda=2\lambda$ is the cosmological constant. This solution is asymptotically of constant curvature $\Lambda$. The parameter $b$ is not related with the mass or the angular momentum of the black hole, so it is considered a hair parameter.\\ 

\begin{figure}[t!]
\centering
\def\svgwidth{0.75\columnwidth}
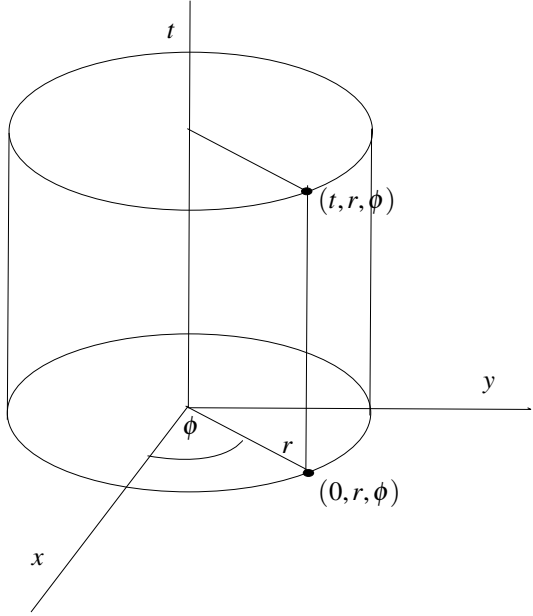
\caption{Coordinates used in the metric in Equation \eqref{metricgen}.}
\label{coordinates}
\end{figure}

From the metric in Equation \eqref{metricgen}, the curvature scalar $R$ is $R=6\Lambda-\dfrac{2b}{r}$. This means that, unlike the BTZ solution, the metric has a curvature singularity at the origin. Depending on the value of the cosmological constant, we have the following kinds of solutions:

\subsection{{Solutions with} $\Lambda<0$}
If we call $\Lambda=-\dfrac{1}{l^2}$ and make the following coordinates transformation 
$\psi\rightarrow i\,t, \, \varphi\rightarrow\phi$ in Equation \eqref{metricgen}, we obtain the solution
\begin{equation}
\label{metricads}
ds^2=-\left(\dfrac{r^2}{l^2}+br-\mu\right)dt^2 +\dfrac{dr^2}{\dfrac{r^2}{l^2}+br-\mu}+r^2 d\phi^2
\end{equation}
for the NMG field equations, where $-\infty<t<+\infty$, $0<\phi< 2\pi$ and $r>0$. In Figure \ref{coordinates} the coordinates used in metric in Equation \eqref{metricads} are shown. It can be seen there are two spacial coordinates, $r$ and $\phi$, and one temporal coordinate, $t$. This coordinate system is used throughout the present article. This solution describes an asymptotically AdS black hole with an inner and an outer event horizon, $r_-$ and $r_+$, such that the metric in Equation \eqref{metricads} can be written as presented in Equation \eqref{metricads2}
\begin{equation}
ds^2=-\dfrac{1}{l^2}(r-r_+)(r-r_-)dt^2 +\dfrac{l^2 dr^2}{(r-r_+)(r-r_-)}+r^2 d\phi^2,
\label{metricads2}
\end{equation}
and the parameters are given by $b=-\frac{1}{l^2}(r_+ +r_-)$ and $\mu=-\frac{r_+ r_-}{l^2}$. In the case of $b=0$, the solution reduces to the static BTZ black hole. It was found the parameter $b$ determines the causal structure of the black hole (\cite{Oliva}). 

\subsection{{Solutions with} $\Lambda>0$}
If we now call $\Lambda=\dfrac{1}{l^2}$, i.e., a positive cosmological constant, and then we make the same type of coordinate transformation $\psi\rightarrow i\,t, \, \varphi\rightarrow\phi$ in Equation \eqref{metricgen}, we obtain the solution
\begin{equation}
\label{metricds}
ds^2=-\left(-\dfrac{r^2}{l^2}+br-\mu\right)dt^2 +\dfrac{dr^2}{-\dfrac{r^2}{l^2}+br-\mu}+r^2 d\phi^2
\end{equation}
which describes an asymptotically \mbox{de Sitter} (dS) black hole. It has two horizons, $r_{++}$ and $r_+$, where $r_+$ is the event horizon which is surrounded by the cosmological horizon  $r_{++}$ (\cite{Oliva}). The metric in Equation \eqref{metricds} can be written as the one presented in Equation \eqref{metricds2}
{\small
\begin{equation*}
ds^2=-\dfrac{1}{l^2}(r-r_+)(r_{++}-r)dt^2 +\dfrac{l^2 dr^2}{(r-r_+)(r_{++}-r)}+r^2 d\phi^2,
\label{metricds2}
\end{equation*}}
and the parameters are given by $b=\frac{1}{l^2}(r_+ +r_{++})$ and $\mu=\frac{r_+ r_{++}}{l^2}$. In this case, the parameter $\mu$ needs to satisfy the following inequality $0<\mu\leq\dfrac{1}{4}b^2 l^2$.

\subsection{{Solutions with} $\Lambda=0$}
NMG admits solutions without cosmological constant and constant curvature. In that case, the metric is given by Equation \eqref{metriclam0}
\begin{equation}
ds^2=-(br-\mu)dt^2 + \dfrac{dr^2}{br-\mu}+r^2 d\phi^2
\label{metriclam0}
\end{equation}
with an event horizon at $r=\dfrac{\mu}{b}$.

\section{{ROTATING BLACK HOLE}}\label{rotating}

A rotating extension of the asymptotically AdS black hole was presented in \cite{Oliva}. For that case, the solution is the metric
\begin{equation}
\label{rotanteconb}
ds^2=-NFdt^2+\dfrac{dr^2}{F}+r^2(d\phi+N^{\phi}dt)^2
\end{equation}
with the functions defined in Equations \eqref{nmg1}, \eqref{nmg2}, \eqref{nmg3}, \eqref{nmg4} and \eqref{nmg5} as follows  
\begin{equation}
N=\left[1-\dfrac{bl^2}{4\sigma}(1-\xi^{-1})\right]^2,
\label{nmg1}
\end{equation}
\begin{equation}
N^{\phi}=-\dfrac{a}{2r^2}(\mu-b\xi^{-1}\sigma),
\label{nmg2}
\end{equation}
{\small
\begin{equation}
F=\dfrac{\sigma^2}{r^2}\left[\dfrac{\sigma^2}{l^2}+\dfrac{b}{2}(1+\xi^{-1})\sigma+\dfrac{b^2 l^2}{16}(1-\xi^{-1})^2-\mu\xi\right],
\label{nmg3}
\end{equation}}
\begin{equation}
\sigma=\left[r^2-\dfrac{\mu}{2}l^2(1-\xi)-\dfrac{b^2 l^4}{16}(1-\xi^{-1})^2\right]^{\frac{1}{2}},
\label{nmg4}
\end{equation}
\begin{equation}
\xi^2=1-\dfrac{a^2}{l^2},
\label{nmg5}
\end{equation}
where the angular momentum is given by $J=Ma$, $M$ is the mass (measured with respect to the zero mass black hole) and the parameter $\mu$ is related with the mass as $\mu=4GM$. In this metric, the angular momentum parameter $a$ satisfies $-l<a<l$. The extreme case for this solution would be when the condition $|a|=l$ is satisfied, but it is not possible to attain this limit with the metric in the form in Equation \eqref{rotanteconb}, given that $\xi=0$ in this case and the terms $\xi^{-1}$ in the metric diverge.\\

Because of our interest in studying the extreme case of a rotating black hole in NMG theory, a change in the parameter $b$ is needed. As proposed in \cite{Giribet}, the parameter $\widehat{b}$ is defined as $\widehat{b}:=b\xi^{-1}$, and the metric in Equation \eqref{rotanteconb} takes the form
\begin{equation}
\label{rotanteconb'}
ds^2=-\widehat{N}\widehat{F}dt^2+\dfrac{dr^2}{\widehat{F}}+r^2(d\phi+\widehat{N}^{\phi}dt)^2
\end{equation}
with the functions defined in Equations \eqref{nmg1b}, \eqref{nmg2b}, \eqref{nmg3b}, \eqref{nmg4b} and \eqref{nmg5b} as follows
\begin{equation}
\widehat{N}=\left[1+\dfrac{\widehat{b}l^2}{4\widehat{\sigma}}(1-\xi)\right]^2,
\label{nmg1b}
\end{equation}
\begin{equation}
\widehat{N}^{\phi}=-\dfrac{a}{2r^2}(\mu-\widehat{b}\widehat{\sigma}),
\label{nmg2b}
\end{equation}
{\small
\begin{equation}
\widehat{F}=\dfrac{\widehat{\sigma}^2}{r^2}\left[\dfrac{\widehat{\sigma}^2}{l^2}+\dfrac{\widehat{b}}{2}(1+\xi)\widehat{\sigma}+\dfrac{\widehat{b}^2 l^2}{16}(1-\xi)^2-\mu\xi\right],
\label{nmg3b}
\end{equation}}
\begin{equation}
\widehat{\sigma}=\left[r^2-\dfrac{\mu}{2}l^2(1-\xi)-\dfrac{\widehat{b}^2 l^4}{16}(1-\xi)^2\right]^{\frac{1}{2}},
\label{nmg4b}
\end{equation}
\begin{equation}
\xi^2=1-\dfrac{a^2}{l^2}.
\label{nmg5b}
\end{equation}
The metric in Equation \eqref{rotanteconb'} is the one presented in \cite{Giribet}, where the rotational parameter satisfies $-l\leq a \leq l$ and the extreme case $|a|=l$ is included.\\

It can be noticed that making the change $\widehat{b}\ra -\widehat{b},\, \widehat{\sigma}\ra-\widehat{\sigma}$, takes the functions to themselves, that is
$\widehat{N}\ra \widehat{N},\, \widehat{N}^\phi\ra \widehat{N}^\phi,\, \widehat{F}\ra \widehat{F}$. Therefore, this change does not present new metrics, and we take only $\sigma\geq0$ (non-negative branch), which is also the right choice for the BTZ case ($b=0$).

\section{{ANALYSIS OF THE NON-EXTREME AND EXTREME ROTATING BLACK HOLE}}\label{analysis}

In this section we analyse some features of the rotating black hole presented in Equation \eqref{rotanteconb'}, as the singularity, the event horizon, the area of some surfaces in the manifold and finally, the distance to the horizon for radial curves, with emphasis in the geometry of the extreme case. Quantities that refer to the extreme case will be denoted with a subscript $e$.
\subsection{{Curvature singularity}}
We calculate the curvature scalar,
\begin{equation}
\begin{array}{rcl}
 R & = & \dfrac{1}{2r\widehat{N}^2}\Big\{ \widehat{N}\big[r^3(\partial_r\widehat{N}^\phi)^2-2r\widehat{F}\partial_r^2\widehat{N} \\
 && -(3r\partial_r\widehat{F}+2\widehat{F})\partial_r\widehat{N}\big] + r\widehat{F}(\partial_r\widehat{N})^2 \\
 && -2\widehat{N}^2\big[r\partial_r^2\widehat{F}+2\partial_r\widehat{F}\big]\Big\}.
\end{array}
\label{curvature-sin}
\end{equation}
Therefore, from Equation \eqref{curvature-sin}, we have a curvature singularity if $\widehat{N}=0$, and this happens when $\widehat{\sigma} = -\frac{\widehat{b}l^2}{4}(1-\xi)$, which means, as $\widehat{\sigma}\geq0$, that $\widehat{b} < 0$, and the singularity is located at 
\begin{equation}
r_s = \frac{l}{\sqrt{8}}(1-\xi)^{\frac{1}{2}}\left[\widehat{b}^2l^2(1-\xi)+4\mu\right]^{\frac{1}{2}}
\label{singularity-r}
\end{equation}
and, from Equation \eqref{singularity-r}, we have as constraint $\mu \geq \mu_s$, where $\mu_s := -\dfrac{\widehat{b}^2l^2}{4}(1-\xi)$. As $\widehat{b}=0$ is the BTZ black hole, then there is no curvature singularity.
\subsection{{Event Horizon}}
As can be seen in Equation \eqref{rotanteconb'}, the event horizon is located where $\widehat{F}=0$, that is, where 
{\small
$
\frac{\widehat{\sigma}^2}{r^2}\left[\frac{\widehat{\sigma}^2}{l^2}+\frac{\widehat{b}}{2}(1+\xi)\widehat{\sigma}+\frac{\widehat{b}^2 l^2}{16}(1-\xi)^2-\mu\xi\right]=0
$}
is satisfied. This is achieved if
\begin{equation}
\label{condraicesb'1}
\widehat{\sigma}=0
\end{equation}
or \\
\begin{equation}
\label{condraicesb'2-a}
\dfrac{\widehat{\sigma}^2}{l^2}+\dfrac{\widehat{b}}{2}(1+\xi)\widehat{\sigma}+\dfrac{\widehat{b}^2 l^2}{16}(1-\xi)^2-\mu\xi=0.
\end{equation}
If Equation \eqref{condraicesb'1} is satisfied, then we have that the following 
$
r_{\widehat{\sigma}}=\dfrac{l}{4}(1-\xi)^{\frac{1}{2}}\left[8\mu+ \widehat{b}^2l^2 (1-\xi)\right]^{\frac{1}{2}}
$
could be the location of the event horizon. On the other hand, the condition in Equation \eqref{condraicesb'2-a} is equivalent to
$
\widehat{\sigma}_{\pm}=\dfrac{l}{4}\left[-{\widehat{b}l}(1+\xi)\pm 2\xi^{\frac{1}{2}}(\widehat{b}^2l^2+4\mu)^{\frac{1}{2}}\right]
$
and so, the event horizon is at 
$
r_{\pm}=\frac{l}{\sqrt{8}}\,(1+\xi)^{\frac{1}{2}}\left|(\widehat{b}^2l^2+4\mu)^{\frac{1}{2}}\mp l\widehat{b}\xi^{\frac{1}{2}}\right|.
$\\

It means there are three possible roots of $\widehat{F}$, but we are interested in the outer horizon and its relation to the curvature singularity, so it is necessary to determine which of the roots is larger. In order to $r_{\pm}\in\mathbb{R}$ we have to require $\mu \geq -\dfrac{\widehat{b}^2l^2}{4}$. This condition is less restrictive than the condition to ensure $r_{\widehat{\sigma}} \in \mathbb{R}$. Now it is necessary to separate the analysis according to the sign of $\widehat{b}$.

\subsubsection{$\widehat{b}=0$}

\begin{figure}[t!]
\centering
\def\svgwidth{0.75\columnwidth}
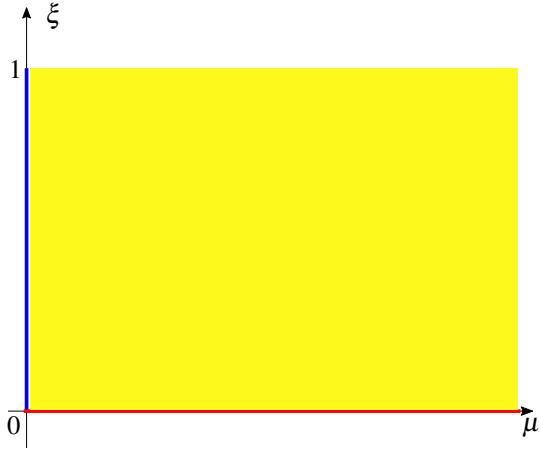
\caption{{Parameter space for $\widehat{b}=0$}: ${(a)}$The blue line represents the massless BTZ black hole, the horizon is a double root of $\widehat{F}$. ${(b)}$The region shaded in yellow represents non-extreme BTZ black holes, the horizon is a simple root of $\widehat{F}$. ${(c)}$The red line represents extreme rotating BTZ black holes, the horizon is a double root of $\widehat{F}$.}
\label{BTZroot}
\end{figure}

If $\widehat{b}=0$, then $r_{\widehat{\sigma}} = \frac{l}{\sqrt{2}}(1-\xi)^{\frac{1}{2}}\mu^{\frac{1}{2}}$ and $r_\pm = \frac{l}{\sqrt{2}}(1+\xi)^{\frac{1}{2}}\mu^{\frac{1}{2}}$, with the constraint $\mu \geq 0$. We have that $r_s=r_{\widehat{\sigma}}$. As $0\leq\xi\leq 1$ then $r_{\widehat{\sigma}} \leq r_\pm$. Also, we have that
$
 \widehat{\sigma}_\pm = \pm l \xi^\frac{1}{2}\mu^\frac{1}{2},
$
which tells us that in general $\widehat{\sigma}_+>0$, $\widehat{\sigma}_-<0$, and that $\widehat{\sigma}_+=\widehat{\sigma}_-$ only if $\mu=0$ or $\xi=0$.\\

So we conclude that the mass needs to satisfy the condition $\mu \geq 0$, that the outer horizon is
$
 r_+ = \frac{l}{\sqrt{2}}(1+\xi)^\frac{1}{2}\mu^\frac{1}{2},
$
and that $\widehat{F}$ has a simple root in $r_+$ unless $\mu=0$ or $\xi=0$, in which case it has a double root. We see that $r_+$ is an increasing function of $\mu$ and also an increasing function of $\xi$. \\

Being more explicit regarding the multiplicity of the roots, if $\xi>0$ and $\mu>0$ then $\widehat{\sigma}_-<0<\widehat{\sigma}_+$ and $r_+>r_{\widehat{\sigma}}\geq0$, which shows that $\widehat{F}$ has a simple root at $r_+$.\\

If $\mu=0$ then the solution does not depend on $\xi$ and $\widehat{F}=\frac{r^2}{l^2}$, which is the massless BTZ black hole.\\

If $\xi=0$ and $\mu>0$ then $\widehat{F}=\frac{\widehat{\sigma}^4}{l^2r^2}$ and $r_+=r_{\widehat{\sigma}}>0$, which shows that $\widehat{F}$ has a double root at $r_+$.\\

These results are summarized in Figure \ref{BTZroot}. Each point in the figure represents a metric with a value of $\mu$ and a value of $\xi$. The allowed values of these parameters are $\mu\geq0$ and $0\leq\xi\leq1$, and the nature of the biggest root of $\widehat{F}$ for each combination of parameters has been made clear by the color coding.


\subsubsection{$\widehat{b}<0$}

If $\widehat{b}<0$ then $r_+ \geq r_-$ and ${\widehat{\sigma}}_+>0$. Therefore we need to compare $r_+$ with $r_{\sigma}$. To have $r_{\sigma}\in\mR$ we need
$
 \mu\geq-\frac{\widehat{b}^2l^2}{8}(1-\xi),
$
which is more restrictive than $\mu\geq-\frac{\widehat{b}^2l^2}{4}$, and we have $r_+ > r_{\widehat{\sigma}}$. Also, we see that $\widehat{\sigma}_+>0$ and that $\widehat{\sigma}_+=\widehat{\sigma}_-$ only if $\mu=-\frac{\widehat{b}^2l^2}{4}$ or $\xi=0$.\\

So we conclude that the mass needs to satisfy the condition
$
 \mu \geq \mu_0,\, \mu_0 := -\frac{\widehat{b}^2l^2}{4},
$
that the outer horizon is
$
 r_+ = \frac{l}{\sqrt{8}}(1+\xi)^\frac{1}{2}\left[(\widehat{b}^2l^2+4\mu)^\frac{1}{2} - \widehat{b}l\xi^\frac{1}{2}\right],
$
and that $\widehat{F}$ has a simple root in $r_+$ unless $\mu=\mu_0$ or $\xi=0$, in which case it has a double root. In general $r_+>0$, and $r_+=0$ only if $\mu=\mu_0$ and $\xi=0$.  We see that $r_+$ is an increasing function of $\mu$ and also an increasing function of $\xi$.\\

Being more explicit regarding the multiplicity of the roots, if $\xi>0$ and $\mu>\mu_0$ then $\widehat{\sigma}_+>\widehat{\sigma}_-$ and $\widehat{\sigma}_+>0$, which shows that $\widehat{F}$ has a simple root at $r_+$.\\

If $\xi>0$ and $\mu=\mu_0$ then $\widehat{\sigma}_+=\widehat{\sigma}_->0$ and $r_+>0$, which shows that $\widehat{F}$ has a double root at $r_+$.\\

If $\xi=0$ and $\mu>\mu_0$ then $\widehat{\sigma}_+=\widehat{\sigma}_->0$ and $r_+>0$, which shows that $\widehat{F}$ has a double root at $r_+$.\\

If $\xi=0$ and $\mu=\mu_0$ then $\widehat{\sigma}_+=\widehat{\sigma}_->0$ and in this case $r_+=0$, which seems to imply that $\widehat{F}$ diverges, but in fact $\widehat{F}$ has a double root at $r_+$, as can be seen by taking a Taylor expansion of $\widehat{\sigma}$ near $r=0$.
\begin{figure}[b!]
\centering
\def\svgwidth{\columnwidth}
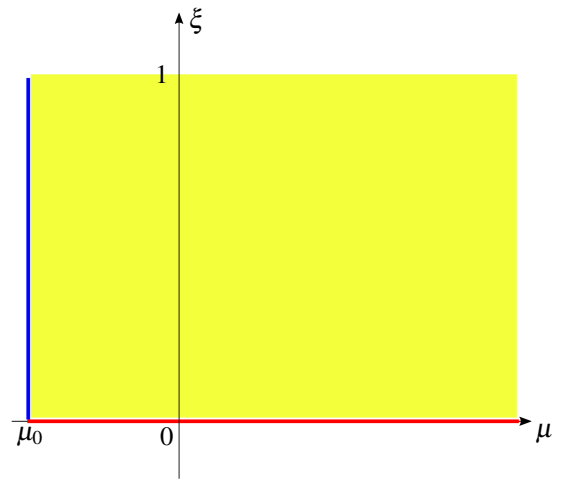
\caption{{Parameter space for $\widehat{b}<0$}: ${(a)}$The blue line represents black holes with minimum mass $\mu_0$, the horizon is a double root of $\widehat{F}$. ${(b)}$The region shaded in yellow represents non-extreme black holes, the horizon is a simple root of $\widehat{F}$. ${(c)}$The red line represents extreme rotating black holes, the horizon is a double root of $\widehat{F}$ and coincides with the curvature singularity.}
\label{b<0root}
\end{figure}
These results are summarized in Figure \ref{b<0root}.\\

With regard to the singularity, if $\mu\geq\mu_s$ then $r_+\geq r_s$. In order for $r_+ = r_s$ we need $\xi = 0$, that is, to be in the extremely rotating case. So in general the singularity is hidden behind the horizon, unless we are in the extremely rotating case, and then the singularity and the event horizon coincide.\\

In the extreme case, $|a|=l$, we have $\xi_e=0$ which can be replaced in the formula above or we can obtain the horizon from  
$
\widehat{F}_e=\frac{\widehat{\sigma}_e^2}{l^2r^2}\left({\widehat{\sigma}_e}+\frac{\widehat{b} l^2}{4}\right)^2,
$ where
$
\widehat{\sigma}_e=\sqrt{r^2-\frac{\mu}{2}l^2-\frac{\widehat{b}^2 l^4}{16}}
$. The condition $\widehat{F}_e=0$ is satisfied if 
\begin{equation}
\label{condsigma'e1}
\widehat{\sigma}_e=0
\end{equation}
or
\begin{equation}
\label{condsigma'e}
\widehat{\sigma}_e=-\dfrac{\widehat{b}l^2}{4}.
\end{equation}
If Equation \eqref{condsigma'e1} is satisfied, we have 
$
r_{\widehat{\sigma}_e}=\dfrac{l}{4}\,\sqrt{8\mu+{\widehat{b}^2 l^2}}
$
while if Equation \eqref{condsigma'e} is true, then $r_{e+}=r_{e-}$ and $r_{e+}$ can be written as
\begin{equation}
\label{horizonteextremo}
r_{e+}=l\sqrt{\dfrac{\Delta\mu}{2}},\qquad \Delta \mu=\mu-\mu_0.
\end{equation}
Given that $\Delta \mu=\mu-\mu_0\geq 0$ must be satisfied, the mass satisfies $\mu\geq \mu_0$. This condition is also presented in \cite{Giribet}. \\

On the other hand, it also can be proven by algebraic calculations that $r_{e+}\geq r_{\widehat{\sigma}_e}$ and so, the outer horizon is given by Equation \eqref{horizonteextremo} in the rotating extreme case. Additionally, one can prove that $r_{e+}\leq r_+$ and the equality is obtained only in the extreme case.

\subsubsection{$\widehat{b}>0$}

\begin{figure}[b!]
\centering
\def\svgwidth{0.90\columnwidth}
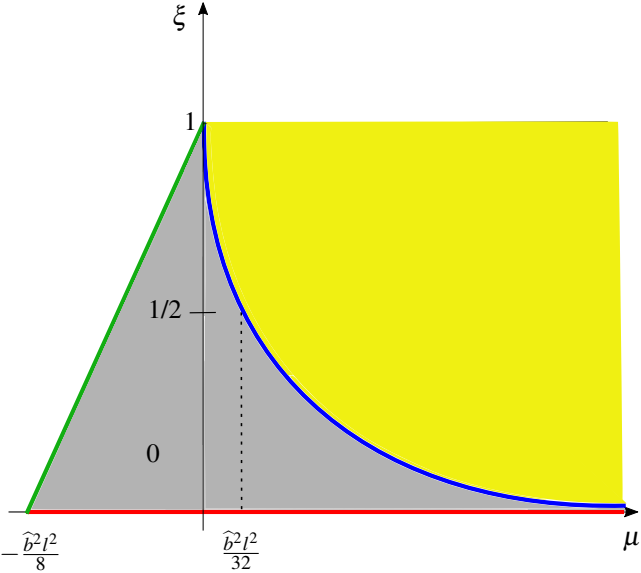
\caption{{Parameter space for $\widehat{b}>0$}: ${(a)}$The blue line represents black holes with mass $\mu_+$, the horizon is a $(r-r_{\widehat{\sigma}})^\frac{3}{2}$ root of $\widehat{F}$. ${(b)}$The region shaded in yellow represents non-extreme black holes, the horizon is a simple root of $\widehat{F}$. ${(c)}$The region shaded in grey represents non-extreme black holes, the horizon is a simple root of $\widehat{F}$. ${(d)}$The red line represents extreme rotating black holes, the horizon is a simple root of $\widehat{F}$. ${(e)}$The green line represents black holes with minimum mass $\mu_-$, $\widehat{F}$ has no zero.}
\label{b>0root}
\end{figure}

If $\widehat{b}>0$ then $\widehat{\sigma}_-<0$. In order for $\widehat{\sigma}_+\geq0$ we need that
$
 \mu\geq \mu_+,\qquad \mu_+ := \frac{\widehat{b}^2l^2}{16}\frac{(1-\xi)^2}{\xi}.
$
Then $r_+\geq r_{\widehat{\sigma}}$ and $r_+ = r_{\widehat{\sigma}}$ only if $\mu = \mu_+$. On the other hand, if
$
 \mu_+\geq\mu\geq \mu_-,\qquad \mu_- := -\frac{\widehat{b}^2l^2}{8}(1-\xi),
$
then $r_{\widehat{\sigma}}\geq0$, and it is the outer horizon. Then $\widehat{\sigma}$ can be written as $ \widehat{\sigma}=(r-r_{\widehat{\sigma}})^{\frac{1}{2}}(r+r_{\widehat{\sigma}})^{\frac{1}{2}}$, but the analysis of the multiplicity of the roots is not as straightforward as before.\\

So we conclude that we have two regions for the mass, if
$
 \mu\geq\mu_+\,\Rightarrow\,r_+=\frac{l}{\sqrt{8}}(1+\xi)^\frac{1}{2}\left[(\widehat{b}^2l^2+4\mu)^\frac{1}{2} - \widehat{b}l\xi^\frac{1}{2}\right],
$
and if
$
 \mu_+\geq\mu\geq\mu_-\,\Rightarrow\, r_+=\frac{l}{4}(1-\xi)^\frac{1}{2}\left[8\mu+\widehat{b}^2l^2(1-\xi)\right]^\frac{1}{2}.
$\\

It can be proved that if $\mu_+\geq\mu\geq\mu_-$, $r_+$ is a decreasing function of $\xi$, in constrast with the case $\widehat{b}<0$. If $\mu\geq\mu_+$, the behaviour of the function $r_+$ is a bit more complicated: if $\mu > \mu_m$, then $r_+$ is an increasing function of $\xi$ while if $\mu<\mu_m$, then $r_+$ is a decreasing function of $\xi$, where $\mu_m=\dfrac{\widehat{b}^2l^2}{4}\left[\dfrac{(1+2\xi)^2}{\xi}-1\right]$.\\

We concentrate now on the multiplicity of the outermost root of $\widehat{F}$. If $\xi>0$ and $\mu>\mu_+$ then $\widehat{\sigma}_+>0>\widehat{\sigma}_-$ and $r_+>r_{\widehat{\sigma}}>0$, then $\widehat{F}$ has a simple root in $r_+$.\\

If $\xi>0$ and $\mu=\mu_+$ then $r_+=r_{\widehat{\sigma}}>0$ and $\widehat{\sigma}_+=0$, $\widehat{\sigma}_-<0$, and therefore we have $\widehat{\sigma}=(r-r_{\widehat{\sigma}})^{\frac{1}{2}}(r+r_{\widehat{\sigma}})^{\frac{1}{2}}$ and
$
 \widehat{F} = (r-r_{\widehat{\sigma}})^\frac{3}{2}\t{F},
$
where
$
 \t{F} = \frac{1}{l^2r^2}(r+r_{\widehat{\sigma}})^\frac{3}{2}\left[(r-r_{\widehat{\sigma}})^\frac{1}{2}(r+r_{\widehat{\sigma}})^\frac{1}{2}-\widehat{\sigma}_-\right].
$
This shows that $\widehat{F}$ has a root that behaves as $(r-r_{\widehat{\sigma}})^\frac{3}{2}$ when $r\rightarrow r_{\widehat{\sigma}}$.\\

If $\mu_-<\mu<\mu_+$ then $0>\widehat{\sigma}_+>\widehat{\sigma}_-$ and $r_{\widehat{\sigma}}>0$, and therefore $\widehat{\sigma}=(r-r_{\widehat{\sigma}})^{\frac{1}{2}}(r+r_{\widehat{\sigma}})^{\frac{1}{2}}$ and $\widehat{F} = (r-r_{\widehat{\sigma}})\bar{F}$, where $\bar{F}$ is defined in Equation \eqref{barF}
\begin{equation}
\begin{array}{rcl}
\bar{F} & = & \frac{1}{l^2r^2}(r+r_{\widehat{\sigma}})\left[(r-r_{\widehat{\sigma}})^\frac{1}{2}(r+r_{\widehat{\sigma}})^\frac{1}{2}-{\widehat{\sigma}}_-\right] \\
 && \times\left[(r-r_{\widehat{\sigma}})^\frac{1}{2}(r+r_{\widehat{\sigma}})^\frac{1}{2}-{\widehat{\sigma}}_+\right].
 \label{barF}
\end{array}
\end{equation}

So $\widehat{F}$ has a simple root at $r_{\widehat{\sigma}}$.\\

If $\mu=\mu_-$ then $0>\widehat{\sigma}_+>\widehat{\sigma}_-$ and $r_{\widehat{\sigma}}=0$, and therefore $\widehat{\sigma}=r$ and $\widehat{F} = \frac{1}{l^2}(r-\widehat{\sigma}_+)(r-\widehat{\sigma}_-)$, and $\widehat{F}$ has no root. These results are summarized in Figure \ref{b>0root}. 

\subsection{{Area of the black hole}}
If we consider the hyper-surface $t=t_0$, where $t_0$ is a constant, we have that the induced metric in the hyper-surface is
\begin{equation}
\label{tconst}
ds^2_{t_0}=\dfrac{dr^2}{\widehat{F}}+r^2 d\phi^2.
\end{equation}
Then, we consider the hyper-surface of $t=t_0$ at $r=r_0$, where $r_0$ is a constant. We have that the induced metric in the hyper-surfaces is
\begin{equation}
\label{rconst}
ds^2_{t_0,r_0}=r_0 ^2 d\phi^2.
\end{equation}
In this hyper-surface, we are interested in calculating the area. From Equation \eqref{rconst}, the area differential is $dA=r_0 d\phi$ and, given that $0<\phi< 2\pi$, the total area is $A=2\pi r_0$, which corresponds to the length of a circumference of radius $r_0$. At the extreme case, the area of the rotating black hole is $A_{e+}=2\pi r_{e+}$. \\

As $r_0\geq r_+$, if we denote by $A_+$ the area of the black hole, we have that $A\geq A_+$, and it is equal if and only if $r_0=r_+$.\\

If we consider the case $\widehat{b}\leq0$ and fix the mass $\mu$, as we have $r_{e+}\leq r_+$, then $A_+\geq A_{e+}$, where $A_{e+}$ is the area of the extreme black hole, and it is equal if and only if $r_+=r_{e+}$, that is in the extreme case. It means the horizon area in the extreme case is the area minimizer for the family of black holes.\\

For the $\widehat{b}>0$ case, given the different behaviours of $r_+$ depending on the values of $\mu$, the area minimizer is different for each case. If $\mu_-<\mu<\mu_+$, then the minimum area corresponds to the black hole with the smallest angular momentum. On the other range of values, where $\mu\geq\mu_+$, the minimum area is obtained as follows: if $0\leq\mu\leq \dfrac{7}{4}\widehat{b}^2l^2$, then it is obtained in the static case; if $\dfrac{7}{4}\widehat{b}^2l^2\leq\mu\leq 2\widehat{b}^2l^2 $, then the area as a function of $\xi$ has two critical points, where 
$
\xi_c=\dfrac{1}{8}\left[\dfrac{\widehat{b}^2l^2}{4}-3-\sqrt{\left(\dfrac{\widehat{b}^2l^2}{4}+1\right)\left(\dfrac{\widehat{b}^2l^2}{4}-7\right)}\right]
$
corresponds to the local minimum for the area; and finally, if $\mu\geq 2\widehat{b}^2l^2$, then the minimum area is in the case where $\xi=\xi_c$. As we have seen, the behaviour of the area in the $\widehat{b}>0$ case is more complicated that the one for the $\widehat{b}<0$ case and the extreme case is never the minimizer of the area of the black hole.  

\subsection{{Distance to the horizon}}
For calculating the radial distance to the horizon, it is necessary to take the hyper-surface $\phi=\phi_0$ where $\phi_0$ is a constant in the hyper-surfaces in Equation \eqref{tconst}. So, the induced metric is
$
ds^2_{t_0,\phi_0}=\frac{dr^2}{\widehat{F}}.
$
Then, the radial distance from a point at $r_1$ to a point at $r_2$ is
$
L=\int_{r_1}^{r_2} {\frac{dr}{\sqrt{\widehat{F}}}}.
$\\

In particular, we want to calculate the distance from the event horizon to an arbitrary point $r_1$ in the non-extreme and extreme case, that is,
$
L_+={\huge \int}_{r_+}^{r_1} {\frac{dr}{\sqrt{\widehat{F}}}}
$ 
and
$
L_e=\int_{r_{e+}}^{r_1} {\frac{dr}{\sqrt{\widehat{F}_e}}}.
$\\

A quick argument allows us to see that for $\widehat{b}\leq 0$, $L_+$ is finite while $L_e$ diverges.\\

As $r_+$ is a simple root of $\widehat{F}$ we have that 
\begin{equation}
L_+\appropto \int_{r_{+}}^{r_1} \dfrac{dr}{\sqrt{r-r_+}}=\int_{0}^{r_1-r_{+}} \dfrac{dr}{\sqrt{r}}=2\sqrt{r_1-r_+}
\label{l+finite}
\end{equation}
which is bounded and Equation \eqref{l+finite} proves that $L_+$ is finite. On the other hand, $r_{e+}$ is a double root of $\widehat{F}_e$, and then
\begin{equation}
L_e\appropto \int_{r_{e+}}^{r_1} \dfrac{dr}{{r-r_+}}= \ln (r-r_{e+})\Big |_{r_{e+}}^{r_1}
\label{le+diverges}
\end{equation}
which is unbounded and Equation \eqref{le+diverges} proves that $L_e$ diverges. This shows that the extreme limit has a cylindrical end.\\

The same argument can be applied to the $b>0$ case, showing that the distance to the horizon is always bounded, therefore not having an extreme limit in the sense of having a cylindrical end.\\

Although lengthy, $L_+$ and $L_e$ can be obtained in closed form, and we do it for $\widehat{b}\leq0$, as it corresponds to the case in which we are most interested. We start noticing that, from the metric in Equation \eqref{rotanteconb'}, $\widehat{F}$ can be written as
\begin{equation}
\label{L_F}
\widehat{F}=\dfrac{\widehat{\sigma}^2}{l^2 r^2}\left[(\widehat{\sigma}+B)^2-C\right]
\end{equation}
where
$
\widehat{\sigma}=(r^2-A)^{\frac{1}{2}},
$
$
A=\frac{l^2}{16}(1-\xi)\left[\widehat{b}^2 l^2(1-\xi)+8\mu\right], 
$
$
B=\frac{\widehat{b}l^2}{4}(1+\xi),
$ and
\begin{equation}
\label{C}
C=\dfrac{l^2}{4}\xi(\widehat{b}^2l^2+4\mu).
\end{equation}
We can notice, from Equation \eqref{C}, that $C\geq 0$, it is $C>0$ in the non-extreme case and $C=0$ in the extreme case. Therefore, the distance between a point at $r_1$ and a point at the outer horizon $r_+$ in the non-extreme case is shown in Equation \eqref{Lsigma}
\begin{equation}
\begin{array}{rcl}
L_+ & = & \int_{r_{+}}^{r_1} \dfrac{dr}{\sqrt{\dfrac{\widehat{\sigma}^2}{l^2 r^2}\left[(\widehat{\sigma}+B)^2-C\right]}} \\
& = & l\ln\left(\dfrac{[(\widehat{\sigma}+B)^2-C]^{\frac{1}{2}}+\widehat{\sigma}+B}{C^{\frac{1}{2}}}\right)\Bigg |_{\widehat{\sigma}_+}^{\widehat{\sigma}_1},
\end{array}
\label{Lsigma}
\end{equation}
\begin{figure}[b!]
\centering
\includegraphics[width=0.8\columnwidth]{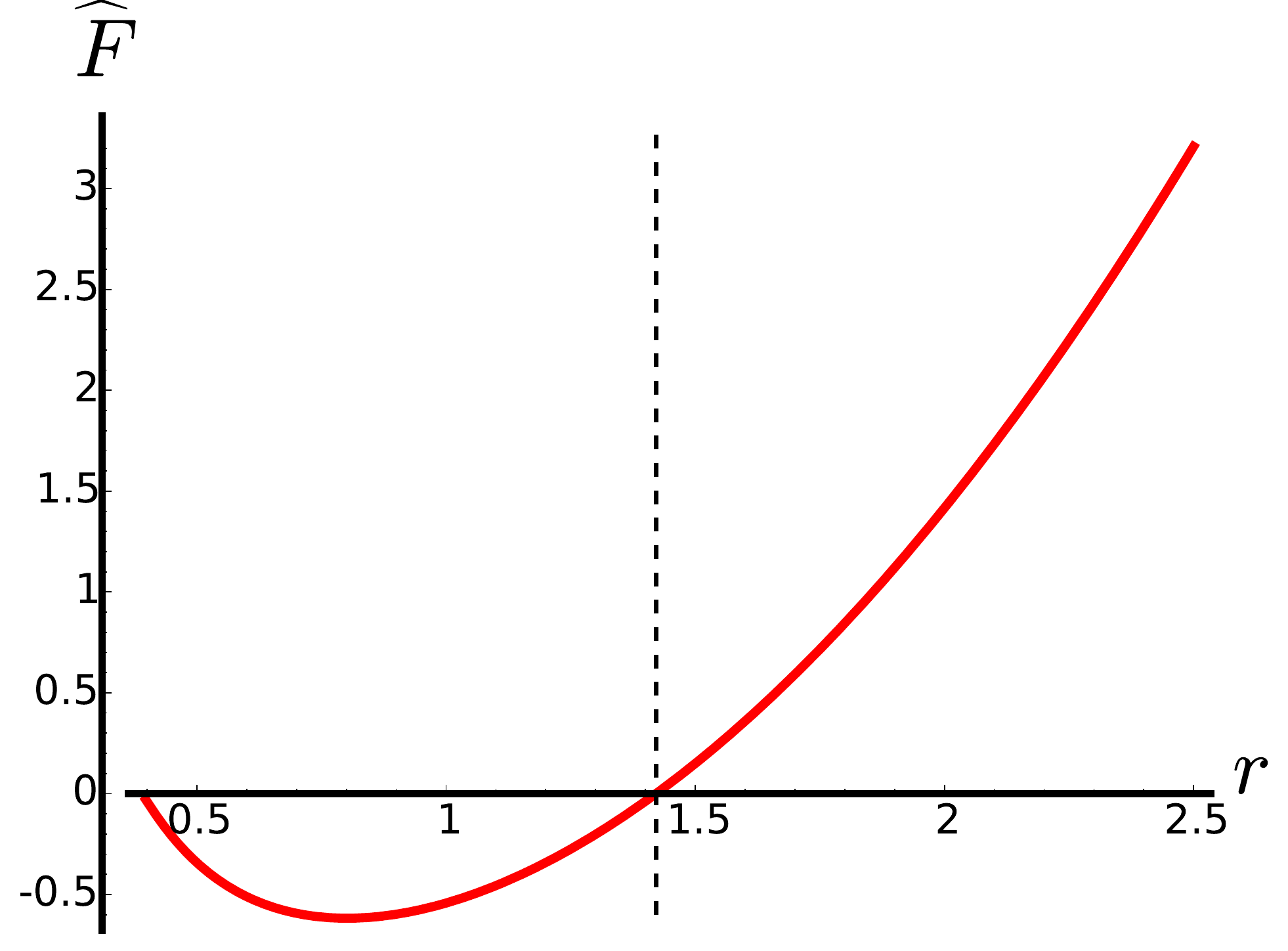}
\caption{$\widehat{F}$ for the non-extreme case with $l=1,\widehat{b}=-1,\mu=1,\xi^2=0.5$. The black dashed line is at the horizon $r_+$.}
\label{Fnoextreme1}
\end{figure}
\begin{figure}[b!]
\centering
\includegraphics[width=0.8\columnwidth]{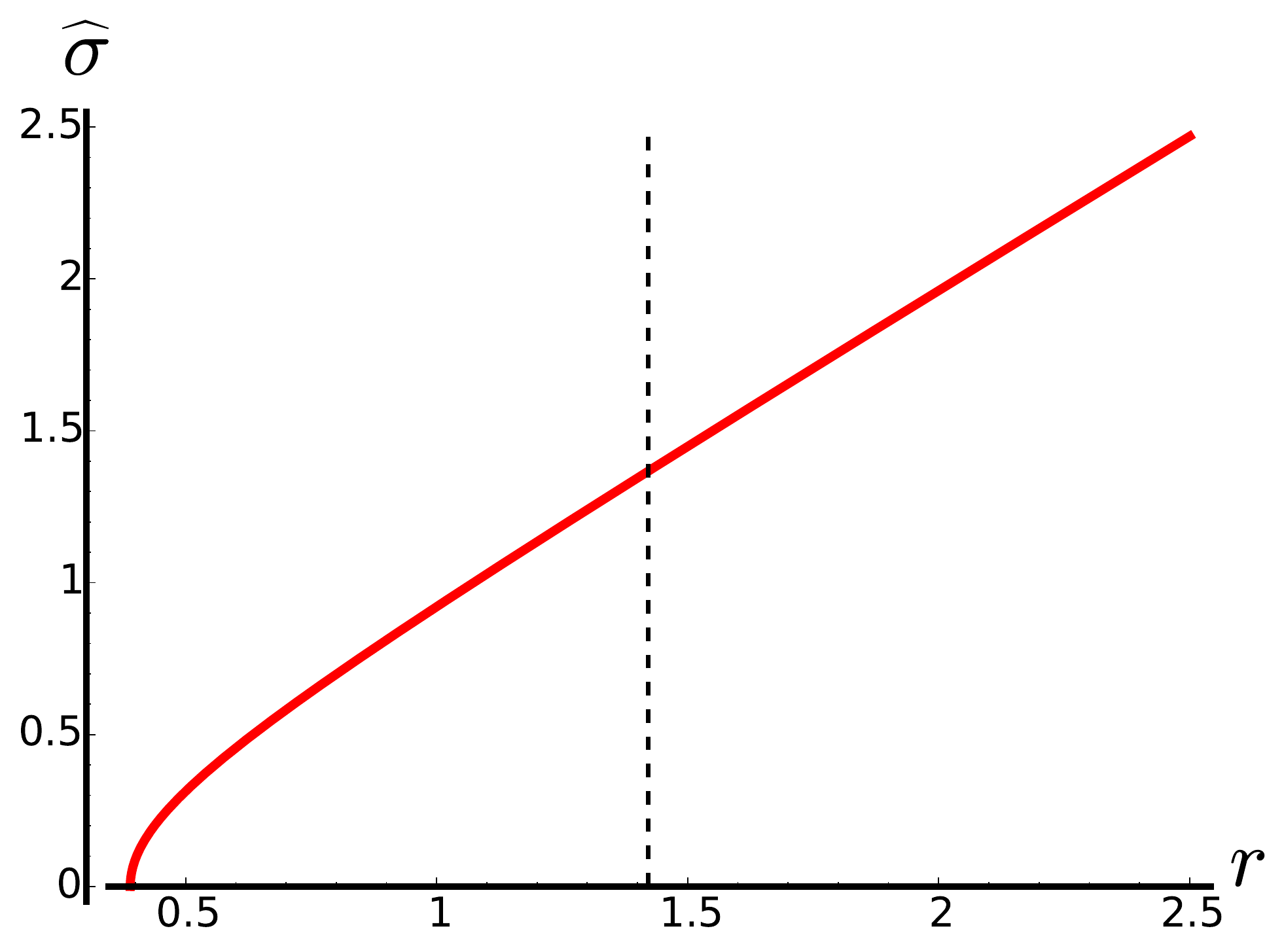}
\caption{$\widehat{\sigma}$ for the non-extreme case with $l=1,\widehat{b}=-1,\mu=1,\xi^2=0.5$. The black dashed line is at the horizon $r_+$.}
\label{snoextreme1}
\end{figure}
where $\widehat{\sigma}_+=(r_+^2-A)^{\frac{1}{2}}$ and $\widehat{\sigma}_1=(r_1^2-A)^{\frac{1}{2}}$. From Equation \eqref{L_F}, this can be written as Equation \eqref{LF}
\begin{equation}
\label{LF}
L_+=l\ln\left(\dfrac{\dfrac{r_1}{\widehat{\sigma}_1}\left(\sqrt{\widehat{F}_1}+\sqrt{\widehat{F}_1+\dfrac{C\widehat{\sigma}_1^2}{l^2r_1^2}}\right)}{\dfrac{r_+}{\widehat{\sigma}_+}\left(\sqrt{\widehat{F}_+}+\sqrt{\widehat{F}_++\dfrac{C\widehat{\sigma}_+^2}{l^2r_+^2}}\right)} \right)
\end{equation}
where $\widehat{F}_1=\widehat{F}(r_1)$ and $\widehat{F}_+=\widehat{F}(r_+)$. Given that $\widehat{F}_+=0$, then $L_+$ is given by Equation \eqref{L}
\begin{equation}
\label{L}
L_+=l\ln\left[\dfrac{lr_1}{C^{\frac{1}{2}}\widehat{\sigma}_1}\left(\sqrt{\widehat{F}_1}+\sqrt{\widehat{F}_1+\dfrac{C\widehat{\sigma}_1^2}{l^2r_1^2}}\right)\right].
\end{equation}
This shows that for the non-extreme case $L_+$ is bounded. We can also see that in the extreme limit, i.e., $C\rightarrow 0$, $L_+$ diverges.\\
 
In Figure \ref{Fnoextreme1} can be seen the plot of $\widehat{F}$ as a function of $r$ for the non-extreme case with $l=1,\,\widehat{b}=-1,\,\mu=1,\,\xi^2=0.5$, while in Figure \ref{snoextreme1} can be seen the plot of $\widehat{\sigma}$ for the same values of the parameters. It can be noted that $\widehat{F}>0$ after the outer horizon, represented by the dashed line, and that $\widehat{\sigma}$ is positive and bounded away from zero for $r>r_+$.\\

\begin{figure}[b!]
\centering
\includegraphics[width=0.8\columnwidth]{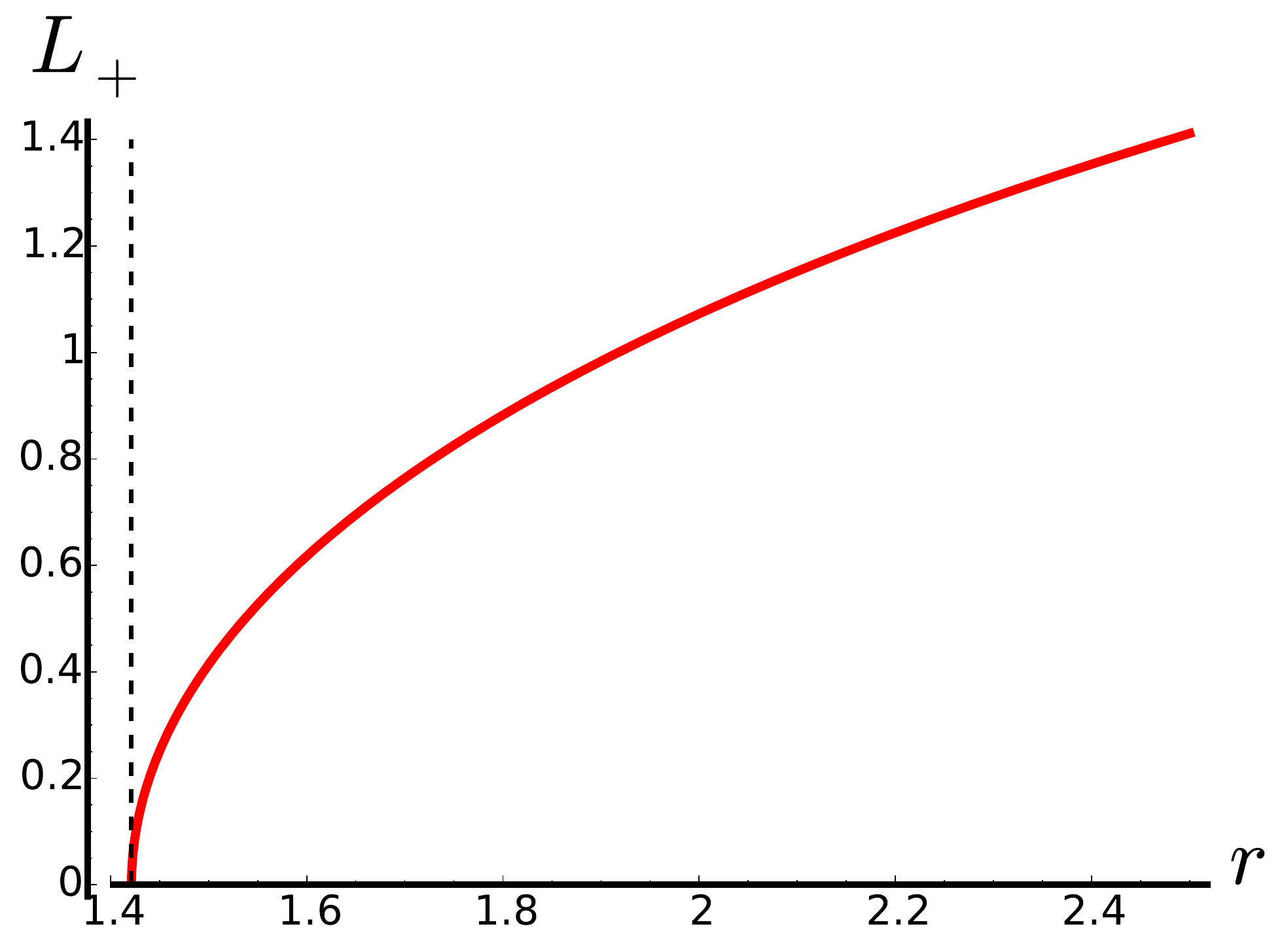}
\caption{Distance to the horizon for the case with $l=1,\widehat{b}=-1,\mu=1,\xi^2=0.5$. The black dashed line is at the horizon $r_+$.}
\label{noextreme1}
\end{figure}
\begin{figure}[b!]
\centering
\includegraphics[width=0.7\columnwidth]{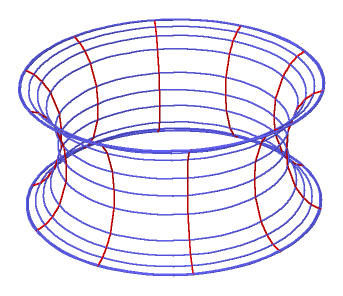}
\caption{Metric structure near the horizon for the non-extreme case, with $l=1,\widehat{b}=-1,\mu=1,\xi^2=0.5$.}
\label{noextreme2}
\end{figure}

In Figure \ref{noextreme1} we present the plot of the distance to the horizon as a function of $r$ for the non-extreme case, again with $l=1,\,\widehat{b}=-1\,,\mu=1,\,\xi^2=0.5$. \\

Also, in Figure \ref{noextreme2} the metric structure near the horizon has been ploted. That is, the plot faithfully represents distances near the horizon. Here the similarity with the structure of the non-extreme Kerr black hole in (3+1)-General Relativity, represented in Figure \ref{kerrnoextr}, is evident. \\

On the other hand, for the extreme case we have
\begin{equation}
\label{Le_F}
\widehat{F}_e=\dfrac{\widehat{\sigma}_e^2}{l^2 r^2}(\widehat{\sigma}_e+B_e)^2
\end{equation}
where
$
\widehat{\sigma}_e=(r^2-A_e)^{\frac{1}{2}},
$
$
A_e=\dfrac{l^2}{16}\left(\widehat{b}^2 l^2+8\mu\right), 
$
$
B_e=\dfrac{\widehat{b}l^2}{4}.
$
\begin{figure}[b!]
\centering
\includegraphics[width=0.8\columnwidth]{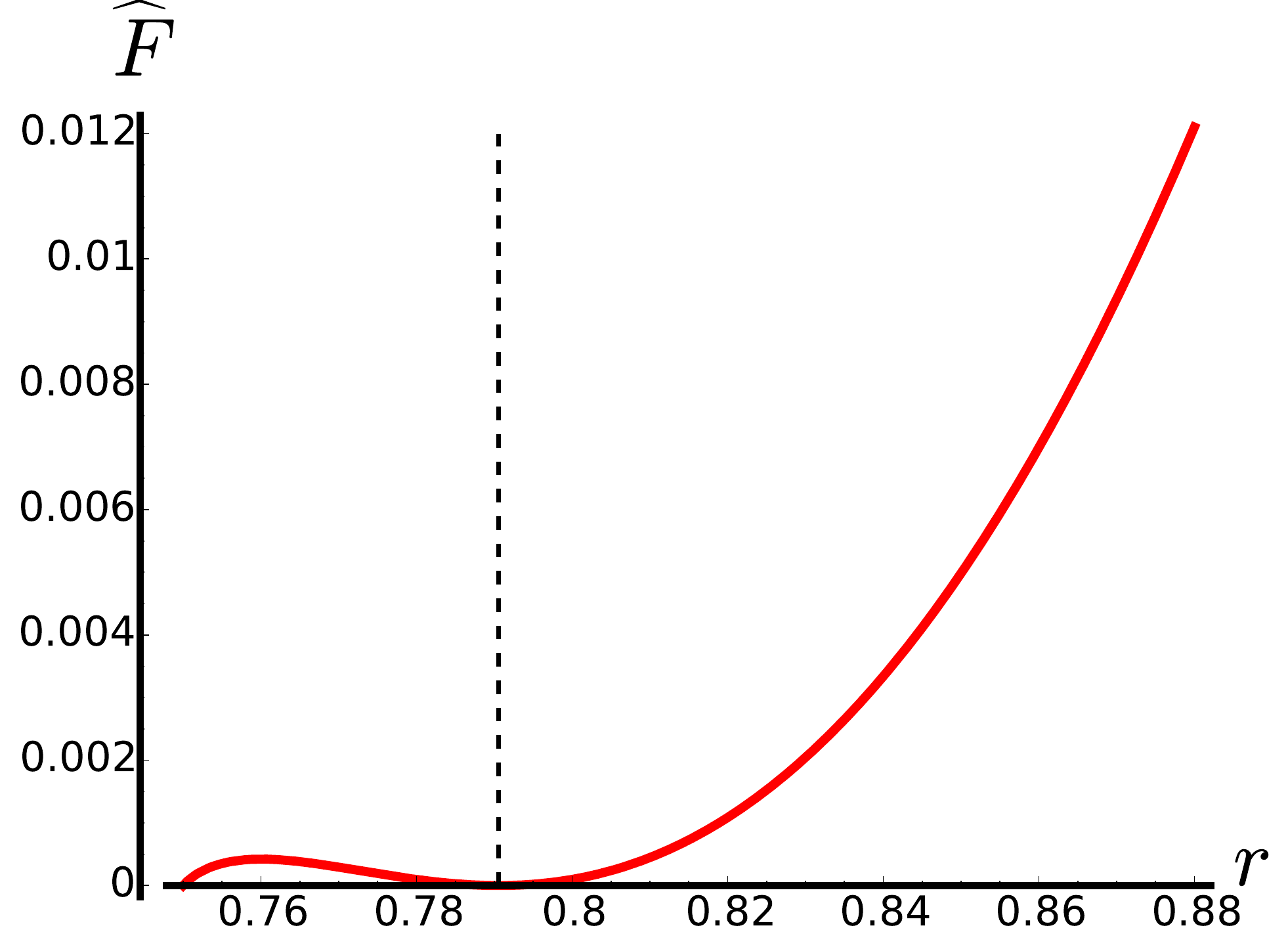}
\caption{$\widehat{F}_e$ for the extreme case with $l=1,\widehat{b}=-1,\mu=1$. The black dashed line is at the horizon $r_{e+}$.}
\label{Fextreme1}
\end{figure}
\begin{figure}[b!]
\centering
\includegraphics[width=0.8\columnwidth]{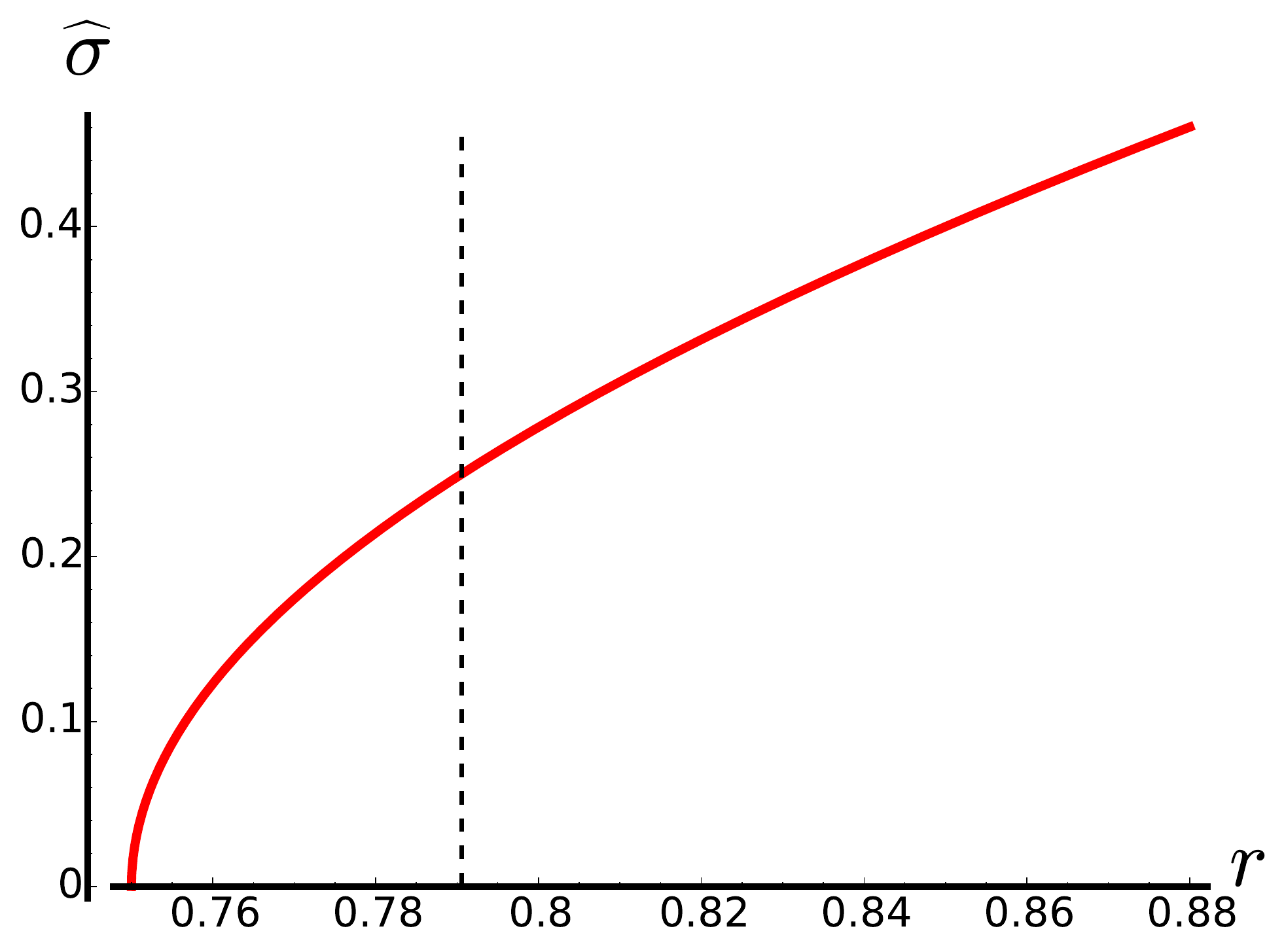}
\caption{$\widehat{\sigma}_e$ for the extreme case with $l=1,\widehat{b}=-1,\mu=1$. The black dashed line is at the horizon $r_{e+}$.}
\label{sextreme1}
\end{figure}
Therefore, the distance between a point at $r_1$ and a point at the outer horizon $r_{e+}$ is given by Equation \eqref{Lesigma}
\begin{equation}
\begin{array}{rcl}
L_e & = & l\int_{r_{e+}}^{r_1} \dfrac{rdr}{{\widehat{\sigma}_e}(\widehat{\sigma}_e+B_e)} \\
& = & l\ln\left(\widehat{\sigma}_e+B_e \right)\Big |_{\widehat{\sigma}_{e+}}^{\widehat{\sigma}_{e1}}
\end{array}
\label{Lesigma} 
\end{equation}
where $\widehat{\sigma}_{e+}=(r_{e+}^2-A_e)^{\frac{1}{2}}$ and $\widehat{\sigma}_{e1}=(r_1^2-A_e)^{\frac{1}{2}}$. From Equation \eqref{Le_F}, this can be written as
\begin{equation}
\label{LeF}
L_e=l\ln\left( \dfrac{r_1}{r_{e+}}  \dfrac{\widehat{\sigma}_{e+}}{\widehat{\sigma}_{e1}}   \dfrac{\sqrt{F_{e1}}}{\sqrt{F_{e+}}}\right)
\end{equation}
where $\widehat{F}_{e1}=\widehat{F}_{e}(r_1)$ and $\widehat{F}_{e+}=\widehat{F}_{e}(r_{e+})$. Given that $\widehat{F}_{e+}=0$, then $L_e$ is unbounded. That is, the outer event horizon in the extreme rotating black hole is located at infinite distance from any point in the hypersurface.\\
\begin{figure}[t!]
\centering
\includegraphics[width=0.8\columnwidth]{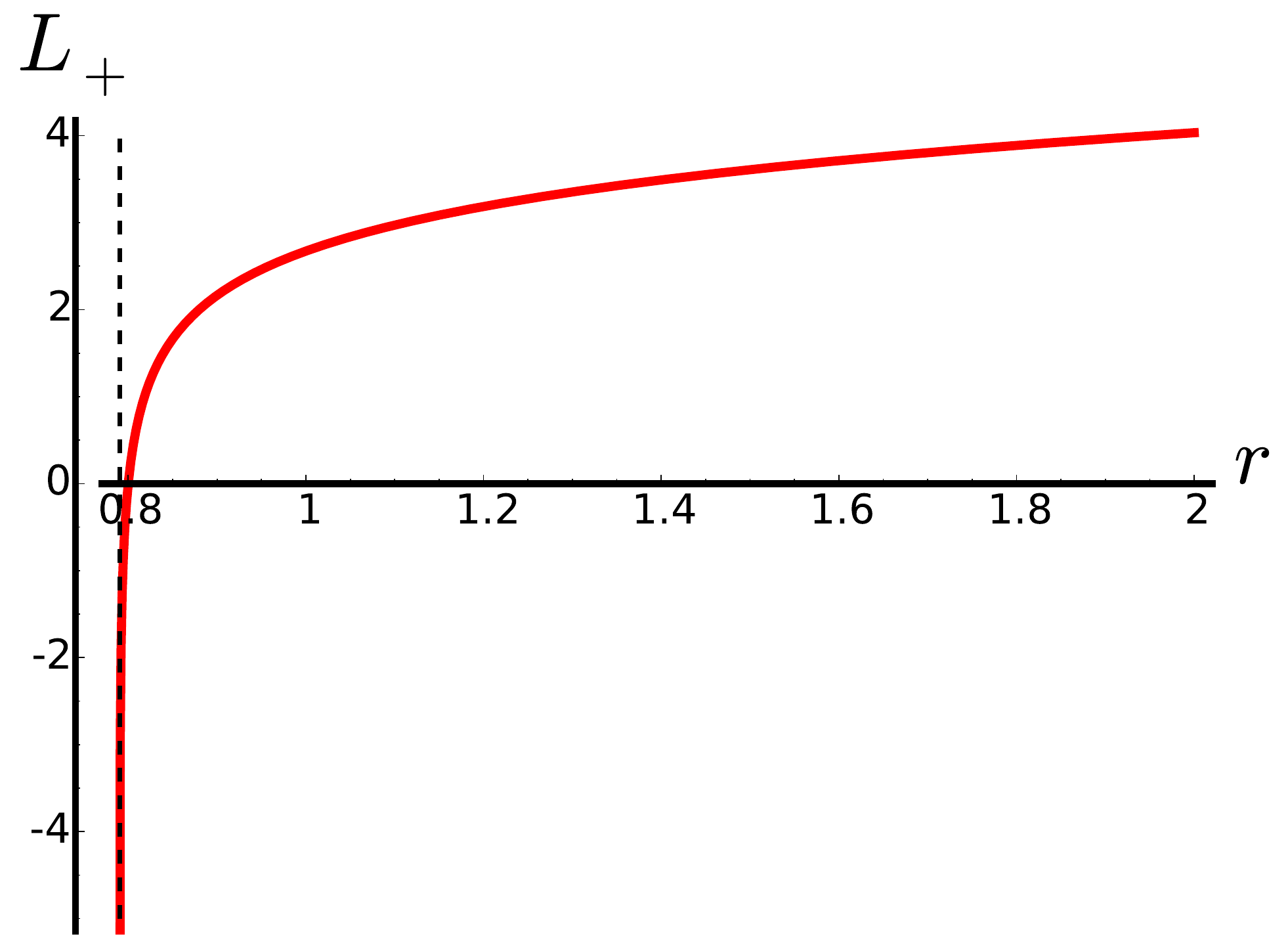}
\caption{Distance to $r=0.8$ for the case with $l=1,\widehat{b}=-1,\mu=1,\xi=0$. The black dashed line is at the horizon $r_{e+}$.}
\label{extreme1}
\end{figure}
\begin{figure}[t!]
\centering
\includegraphics[width=0.7\columnwidth]{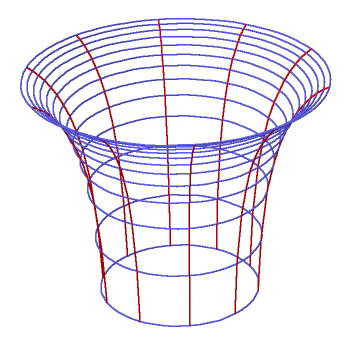}
\caption{Metric structure near the horizon for the extreme case, with $l=1,\widehat{b}=-1,\mu=1,\xi=0$.}
\label{extreme2}
\end{figure}

In Figure \ref{Fextreme1} can be seen the plot of $\widehat{F}_{e}$ as a function of $r$ for the extreme case with $l=1,\,\widehat{b}=-1,\,\mu=1$, while in Figure \ref{sextreme1} can be seen the plot of $\widehat{\sigma}_e$ for the same values of the parameters. It can be noticed that as in the non-extreme case, $\widehat{F}_e>0$ after the outer horizon, represented by the dashed line, and that $\widehat{\sigma}_{e}$ is positive and bounded away from zero for $r>r_{e+}$.\\

In Figure \ref{extreme1} we present the plot of the radial distance to the arbitrarily chosen point $r=0.8$ as a function of $r$ for the extreme case with the same values of the parameters as before, namely, $l=1,\,\widehat{b}=-1\,,\mu=1$, and we see how the distance diverges at the horizon. Also, in Figure \ref{extreme2} the metric structure near the horizon has been ploted. As in the non-extreme case, this is a faithfull representation of distances near the horizon, and it can be seen directly the similarity with the structure of the extreme Kerr black hole in (3+1)-General Relativity, represented in Figure \ref{kerrextr}.

\section{{CONCLUSIONS}}\label{conclusions}

The different cases of a rotating black hole in NMG were analysed. This family of black holes has several interesting features and resembles closely black holes in (3+1)-General Relativity. In particular, the black hole can have several horizons, or none and then it describes a naked singularity.\\

The singularity and the horizon were analysed, showing different features depending on the sign of the hair parameter $\widehat{b}$.\\

Of particular interest to us is the case $\widehat{b}\leq0$, in the sense that it has an extreme limit with respect to the angular momentum parameter. We calculated the distance to the horizon in the non-extreme and extreme cases, showing that in the non-extreme case it is finite while in the extreme case it diverges. This shows that the initial data for the black hole changes structure from black hole initial data to trumpet initial data. This same change occurs in the Kerr spacetime in (3+1)-General Relativity. \\

Also, the area of the horizon was calculated, showing that, being the other parameters fixed and $\widehat{b}\leq0$, the minimimum of the area occurs for the extreme case. This again coincides with what happens in the Kerr spacetime.\\

These characteristics make this black hole a suitable candidate for conjecturing geometrical inequalities in NMG, playing the role that Kerr had for geometrical inequalities in (3+1)-dimensions.

\section{{ACKNOWLEDGMENT}}

We would like to acknowledge the use of SageManifolds (\cite{sagemanifolds}) to perform some of the calculations presented. \\

Andr\'es Ace\~na acknowledges the support of SENESCYT (Ecuador) through a Prometeo grant. 

\begin{center}
\bf REFERENCES
\end{center}

\begingroup
\renewcommand{\section}[2]{}

\setlength{\bibhang}{0pt} 

\endgroup  

\vspace{1cm}

\begin{wrapfigure}{l}{0.20\columnwidth}
\includegraphics[width=0.20\columnwidth]{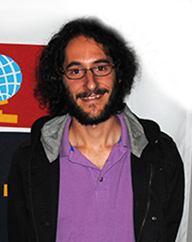}
\end{wrapfigure}\textbf{Andrés Aceña}. Licenciado en Física graduado en la Universidad Nacional de Córdoba.Obtuvo su doctorado en Física Teórica de la Universidad de Postdam/Max Planck Institute for Gravitational Physics, Alemania. Es investigador asistente de la Carrera de Investigador Científico y Tecnológico de CONICET, Argentina. 
Es profesor adjunto de la Facultad de Ciencias Exactas y Naturales de la Universidad Nacional de Cuyo, Argentina. Su área de investigación es la Relatividad General, siendo sus intereses principales de investigación: las desigualdades geométricas, las soluciones en dimensiones distintas a 4, los espacio-tiempos estacionarios asintóticamente planos y la estructura asintótica de las soluciones de las ecuaciones de Einstein. Fue investigador Prometeo del Observatorio Astronómico de Quito y actualmente colabora en el Grupo de Gravitación y Cosmología de esta institución.

\begin{wrapfigure}{l}{0.20\columnwidth}
\includegraphics[width=0.20\columnwidth]{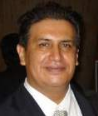}
\end{wrapfigure}\textbf{Ericson López}. Doctor en Astrofísica (PhD) por la   Academia de Ciencias de Rusia y Físico Teórico por la Escuela  Politécnica  Nacional. Ha realizado investigaciones post doctorales en Brasil y Estados Unidos. Es científico colaborador   del   Harvard-Smithsonian   Center para   la   Astrofísica y profesor adjunto del Departamento  de  Astronomía  de  la  Universidad  de  Sao  Paulo.  Ha realizado más   de   30   publicaciones   científicas   y   varias   otras publicaciones relevantes. Es Director del Observatorio Astronómico de  Quito  desde  1997  y  miembro  de  la  Academia  de  Ciencias  del Ecuador. Es profesor principal de la Facultad de Ciencias de la EPN por más de 25 años,  en la que imparte  cursos formales de Física  y Astrofísica.

\begin{wrapfigure}{l}{0.20\columnwidth}
\includegraphics[width=0.20\columnwidth]{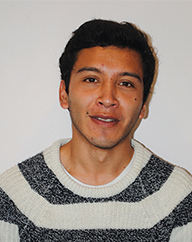}
\end{wrapfigure}\textbf{Mario Llerena}. Realizó sus estudios de pregrado en la Escuela Politécnica Nacional, obteniendo su título   de   Físico   en   el   2016.  Es   miembro   del Observatorio  Astronómico  de  Quito  como  parte de   Unidades   Científicas   de   investigación   en Gravitación   y   Cosmología,   Radioastronomía   y Astrofísica de Altas Energías.

\label{UltimaPagina}


\end{document}